\documentclass[runningheads]{llncs}
\usepackage[T1]{fontenc}
\usepackage{graphicx}
\usepackage{amsmath}
\usepackage{amssymb}
\usepackage{booktabs}
\usepackage{multirow}
\usepackage[table]{xcolor}
\usepackage{hyperref}
\definecolor{oursrow}{RGB}{220,230,241} 
\usepackage{color}
\usepackage{booktabs}  
\begin{document}
\title{TuLaBM: Tumor-Biased Latent Bridge Matching for Contrast-Enhanced MRI Synthesis}

\titlerunning{TuLaBM for Contrast-Enhanced MRI Synthesis}
\author{
Atharva Rege$^{1,2}$, Adinath Madhavrao Dukre$^{1}$, Numan Balci$^{3}$,  \\ Dwarikanath Mahapatra$^{4}$, Imran Razzak$^{1}$ }
 \institute{
$^{1}$MBZUAI, Abu Dhabi, UAE \\
$^{2}$National Institute of Technology Karnataka, Surathkal, India \\
$^{3}$Cleveland Clinic Abu Dhabi, UAE\\
$^{4}$Khalifa University, Abu Dhabi, UAE\\
}

\maketitle              

\begin{abstract}
Contrast-enhanced magnetic resonance imaging (CE-MRI) plays a crucial role in brain tumor assessment; however, its acquisition requires gadolinium-based contrast agents (GBCAs), which increase costs and raise safety concerns. Consequently, synthesizing CE-MRI from non-contrast MRI (NC-MRI) has emerged as a promising alternative. Early Generative Adversarial Network (GAN)-based approaches suffered from instability and mode collapse, while diffusion models, despite impressive synthesis quality, remain computationally expensive and often fail to faithfully reproduce critical tumor contrast patterns.
To address these limitations, we propose Tumor-Biased Latent Bridge Matching (TuLaBM), which formulates NC-to-CE MRI translation as Brownian bridge transport between source and target distributions in a learned latent space, enabling efficient training and inference. To enhance tumor-region fidelity, we introduce a Tumor-Biased Attention Mechanism (TuBAM) that amplifies tumor-relevant latent features during bridge evolution, along with a boundary-aware loss that constrains tumor interfaces to improve margin sharpness. While bridge matching has been explored for medical image translation in 
pixel space, our latent formulation substantially reduces computational 
cost and inference time.
Experiments on BraTS2023-GLI (BraSyn) and Cleveland Clinic (in-house) liver MRI dataset show that TuLaBM consistently outperforms state-of-the-art baselines on both whole-image and tumor-region metrics, generalizes effectively to unseen liver MRI data in zero-shot and fine-tuned settings, and achieves inference times under 0.097 seconds per image. 
.
\keywords{Contrast-enhanced MRI \and Latent Bridge Matching \and 
Tumor-Biased Attention \and Image Synthesis.}
\end{abstract}

\section{Introduction}

Gadolinium-based contrast agents (GBCAs) play a critical role in magnetic resonance imaging (MRI) by enhancing vascular structures and pathological tissues \cite{zhou2013gadolinium}, thus substantially improving tumor detection. Despite their clinical utility, the use of GBCAs is associated with notable drawbacks, including increased scanning costs \cite{xu2021synthesis}, potential adverse effects in patients with chronic kidney disease \cite{schieda2018gadolinium,woolen2020risk}, long-term gadolinium deposition in the brain \cite{gulani2017gadolinium}, and broader environmental concerns \cite{brunjes2020anthropogenic,inoue2020impact}. These limitations have spurred an increasing interest in approaches that reduce or eliminate the dependence on contrast agents while maintaining similar imaging enhancement for accurate tumor detection.

For this purpose, cross-modality MRI synthesis models have been developed to generate contrast-enhanced MRI from non-contrast MRI \cite{dayarathna2024deep}. Early generative approaches were predominantly based on generative adversarial networks (GANs) \cite{li2024controlnet++}, such as Pix2Pix \cite{isola2017image}. however, these approaches suffered from mode collapse or gradient vanishing. ResViT \cite{dalmaz2022resvit} proposed an approach using vision transformers for MRI synthesis. More recently, diffusion models have gained prominence for medical image synthesis due to their improved generative fidelity \cite{meng2024multi,wang2024mutual,xu2024common,zhou2024cascaded}. Palette \cite{saharia2022palette} applies a standard conditional diffusion process for image-to-image translation. But, these models rely on multi-step reverse processes, resulting in high computational cost and prolonged inference times that hinder clinical deployment. I$^2$SB \cite{liu20232} proposes a class of conditional diffusion models that directly learn the nonlinear diffusion processes between two given distributions, that is, build diffusion bridges between the input and output data distributions in their pixel-spaces. D$^3$M \cite{pang2025d} introduces a  deformation-driven diffusion model for CE-MRI synthesis.

Although these models achieve high-quality CE-MRI synthesis, they operate in pixel-space, which leads to longer training and inference times. Moreover, despite architectural advances, CE-MRI synthesis from NC-MRI remains inherently ill-posed due to the lack of contrast cues in NC-MRI \cite{gui2024cavm}. This leads to existing methods continuing to suffer from false positive and false negative enhancement, particularly in tumor regions with complex and heterogeneous morphology, where precise contrast delineation is clinically critical. Furthermore, existing diffusion-based synthesis models predominantly optimize global reconstruction objectives, which are dominated by large background regions in the MRI. This imbalance suppresses learning signals from small yet clinically critical tumor structures. None of these approaches explicitly bias the generative process toward tumor regions during the learning process itself. In this paper, we present TuLaBM, a framework that formulates NC-to-CE MRI synthesis as a transport problem, with an objective to learn a mapping that transfers samples from the source NC-MRI to the target CE-MRI latent distribution, for efficient CE-MRI synthesis from given NC-MRI. We summarize our main contributions as follows:
\begin{itemize}
    \item[$\bullet$] We present \textbf{Tumor-Biased Latent Bridge Matching (TuLaBM)}, a bridge matching framework that formulates NC-to-CE MRI synthesis as a Brownian bridge between source and target distributions in their latent spaces, enabling efficient CE-MRI synthesis.
    
    \item[$\bullet$] We introduce a \textbf{Tumor-Biased Attention Mechanism (TuBAM)} that selectively amplifies tumor-related latent features in the denoiser model during bridge evolution, and a \textbf{boundary-aware loss} that explicitly constrains tumor interfaces, leading to sharper tumor margins.
    
    \item[$\bullet$] We demonstrate through experiments on \textbf{BraTS2023-GLI (BraSyn)} and \textbf{Cleveland Clinic}-\textbf{In-house Liver MRI} datasets that TuLaBM consistently outperforms state-of-the-art baselines, while substantially reducing inference cost.
\end{itemize}

\section{Methodology}

\subsection{Preliminaries: Bridge Matching}

Diffusion models \cite{ho2020denoising,rombach2022high} have shown remarkable success in image-to-image translation, but are limited by their dependence on a fixed Gaussian prior. More recently, the bridge matching framework \cite{albergo2023stochastic,shi2023diffusion} was introduced, which does not involve any noising mechanism and can be applied to any pair of distributions. 

For any two probability distributions $\pi_0$ and $\pi_1$, we assume access to paired samples $(x_0, x_1) \sim \pi_0 \times \pi_1$, such that $x_0 \sim \pi_0$ and $x_1 \sim \pi_1$ are samples obtained from the same subject and anatomical location. Bridge Matching aims to learn a stochastic trajectory between $\pi_0$ and $\pi_1$, defined by a stochastic differential equation (SDE):

\begin{equation}
\mathrm{d}x_t = \frac{x_1 - x_t}{1 - t}\,\,\mathrm{d}t + \sigma\,\mathrm{d}B_t,
\label{eq:sde}
\end{equation}

where $x_t$ is a stochastic interpolant for a Brownian bridge, $v(x_t,t) = \frac{x_1 - x_t}{1 - t}$ is the velocity (drift) field, and $\sigma$ is the noise parameter that controls the noise magnitude, with $\sigma\,\mathrm{d}B_t$ adding stochastic perturbation. The model is trained to predict the drift field $v_{\theta}$ using the following training objective:
\begin{equation}
\mathcal{L}_{\text{bridge}}
=
\mathbb{E}_{t, x_0, x_1}
\left[
\left\|
\frac{x_1 - x_t}{1 - t}
-
v_\theta(x_t, t)
\right\|^2
\right].
\label{eq:bridge_loss}
\end{equation}
 Predicted drift is used to sample from distribution $\pi_1$ using samples from $\pi_0$.
\begin{figure}[t]
\centering
\includegraphics[width=1\textwidth]{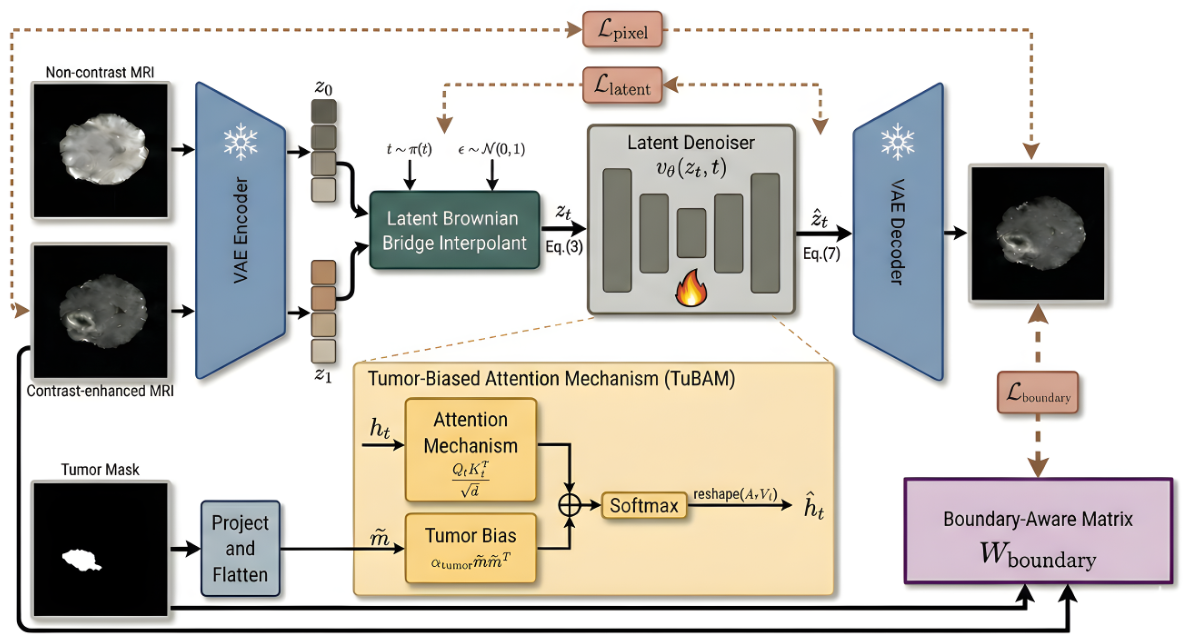}
\caption{Overview of the network architecture of TuLaBM.}
\label{fig:Overview}
\end{figure}

\subsection{Tumor-Biased Latent Bridge Matching}
Directly applying bridge matching in pixel space is computationally expensive and time consuming. Hence, bridge matching is performed within a learned latent space \cite{chadebec2025lbm}. Let $\mathcal{E}(\cdot)$ and $\mathcal{D}(\cdot)$ denote the encoder and decoder of a  pre-trained variational autoencoder (VAE) \cite{kingma2013auto,ronneberger2015u}. Fig.~\ref{fig:Overview} illustrates the overall architecture.

Given paired MRI slices $(x_0, x_1)$, where $x_0$ and $x_1$ represent NC-MRI and CE-MRI respectively. We obtain their latent representations
$z_0 = \mathcal{E}(x_0)$ and $z_1 = \mathcal{E}(x_1)$, where $z_0, z_1 \in \mathcal{Z}$ lie in a lower-dimensional latent space that preserves the global anatomical structure. A latent Brownian bridge interpolant $z_t$ is defined in the latent space $\mathcal{Z}$ as,
\begin{equation}
z_t = (1 - t)\,z_0 + t\,z_1 + \sigma \sqrt{t(1 - t)}\,\epsilon ,
\label{eq:latent_interpolant}
\end{equation}
where $t \in [0,1] $, $\sigma \geq 0$ and $\epsilon \sim \mathcal{N}(0, I)$. A latent drift field $v_\theta(z_t, t)$ that approximates the velocity field of the latent bridge is learned using a denoiser model with training objective similar to Eq.\eqref {eq:bridge_loss} but in latent space:
\begin{equation}
\mathcal{L}_{\mathrm{latent}} =
\mathbb{E}_{t, z_0, z_1}\left[
\left\|
\frac{z_1 - z_t}{1 - t}
- v_\theta(z_t, t)
\right\|^2
\right].
\label{eq:latent_loss}
\end{equation}

\subsubsection{Tumor-Biased Attention Mechanism (TuBAM).}

To explicitly emphasize the tumor region generation during latent transport, we introduce the
Tumor-Biased Attention Mechanism (TuBAM) that injects a structured additive bias into the self-attention logits during training. Let $h_t \in \mathbb{R}^{C \times H_\ell \times W_\ell}$ denote the latent feature map at diffusion timestep $t$. $h_{t}$ is flattened into a sequence of tokens $X_t \in \mathbb{R}^{N \times C}$,  where $N = H_\ell W_\ell$ and each token corresponds to a spatial latent location. Let $m_{tumor} \in \{0,1\}^{H \times W}$ be the ground-truth tumor mask of the input NC-MRI image, available during training. The mask is first projected to the latent resolution $m_\ell = Downsample(m_{tumor}) \in \{0,1\}^{H_\ell \times W_\ell}$ and flattened to align with the attention tokens, to get $\tilde{m} \in \{0,1\}^{N}$.

Latent features are projected into query, key, and value embeddings
$Q_t$,
$K_t$, and
$V_t$. We incorporate tumor awareness by adding an additive bias to the attention logits as follows:
\begin{equation}
A_t^{\text{TuBAM}} = \mathrm{softmax}\!\left(
\frac{Q_t K_t^\top}{\sqrt{d}} + \alpha_{tumor}\, \tilde{m} \tilde{m}^\top
\right),
\end{equation}
where $\alpha_{tumor}$ controls the strength of the tumor bias and the bias term $\tilde{m}\tilde{m}^\top$ selectively reinforces interactions between tumor-associated tokens as,

\begin{equation}
    \tilde{m} \tilde{m}^\top_{ij} =
\begin{cases}
1 & \text{if both tokens } i,j \text{ correspond to tumor regions}, \\
0 & \text{otherwise}.
\end{cases}
\end{equation}
\begin{figure}[!htb]
    \centering
    \includegraphics[width=1\textwidth]{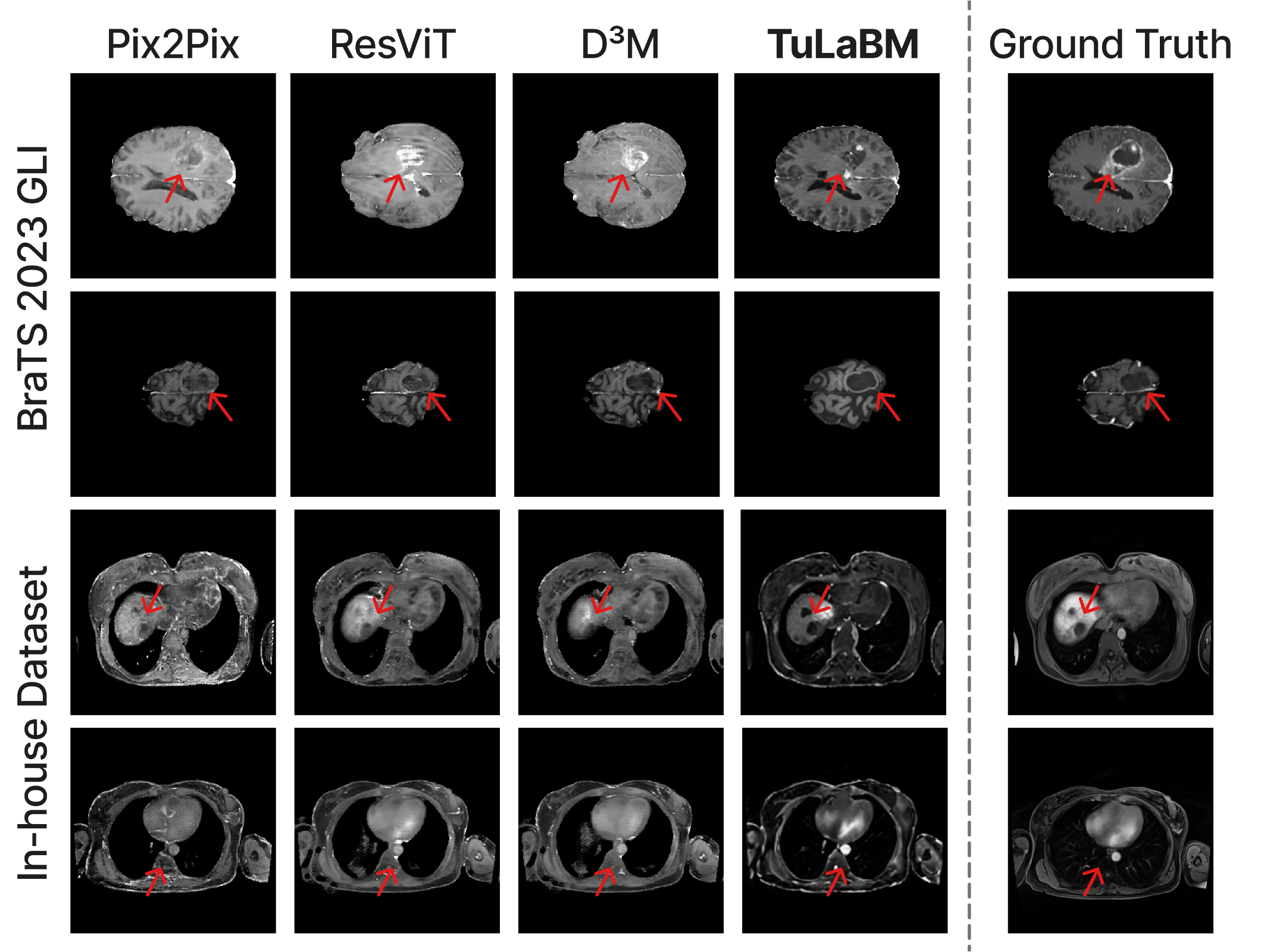}
    \caption{Qualitative comparison of synthesized CE-MRI across methods on the 
    BraTS 2023 GLI (rows 1--2) and in-house liver MRI (rows 3--4) datasets.}
    \label{fig:qualitative}
\end{figure}

The attended latent features are then given by $\hat{h}_t = \mathrm{reshape}(A^{TuBAM}_t . V_t)$, reshaped to the spatial dimensions of the latent feature map. By injecting tumor-aware bias directly into the attention logits, the model encourages stronger intra-tumor feature aggregation during training, increasing the representational emphasis on tumor regions while preserving the normalization and stability of the attention operator. This mechanism is applied only during training to improve the internal representation of tumor-related features during training and does not act as an explicit conditioning input. \\
\subsubsection{Training.} For a timestep $t$ sampled from a timestep distribution $\pi(t)$, $z_t$ is retrieved using Eq.\eqref{eq:latent_interpolant} and passed to the latent denoiser to predict the drift $v_{\theta}(z_t,t)$. The predicted terminal latent $\hat{z}_1$ is computed using the predicted drift as follows: 
\begin{equation}
\hat{z}_1 = (1-t) \cdot v_{\theta}(z_t,t) + z_t .
\label{eq:predict_latent}
\end{equation}

The latent denoiser is trained to regress the latent drift field using the latent loss defined in Eq.\eqref{eq:latent_loss}. To ensure faithful reconstruction in image space, $\hat{z}_1$ is decoded through the VAE decoder $\mathcal{D}(\cdot)$ and compared with the ground-truth CE-MRI $x_1$ using a pixel-wise loss,
\begin{equation}
\mathcal{L}_{\text{pixel}} = \ell(\mathcal{D}(\hat{z}_1), x_1),
\end{equation}
where $\ell(\cdot)$ is $\ell_1$ or $\ell_2$ reconstruction loss.

\paragraph{Boundary-Aware Loss.} To further enhance tumor boundary sharpness, we incorporate a \emph{boundary-aware loss} that explicitly supervises tumor interfaces. For this purpose, we compute a distance-to-boundary map $D_{\text{boundary}} \in [0,1]^{H \times W}$, which encodes the normalized Euclidean distance of each pixel to the nearest tumor boundary in the corresponding mask. Distances are clipped to a maximum radius $d_{\max}$, normalized to $[0,1]$, and pixels outside the tumor region are assigned zero weight. Pixel-wise weights are defined as an exponentially decaying function of the distance-to-boundary map:
\begin{equation}
w(p) = \exp\left(-\frac{D_{\text{boundary}}(p)}{\tau}\right) \cdot m_{\text{tumor}}(p),
\end{equation}
for every pixel p in the image, where $\tau$ is a temperature hyperparameter controlling boundary sharpness, and $m_{\text{tumor}}$ is the binary tumor mask that ensures supervision is restricted to tumor regions. Let us define $W_{\text{boundary}}(p) = w(p)$, for every pixel p. The boundary-aware pixel loss between the target CE-MRI and synthesized output $\mathcal{D}(\hat{z}_1)$ is defined as,
\begin{equation}
\mathcal{L}_{\text{boundary}}
=
\ell\!\left(
W_{\text{boundary}} \odot \hat{x}_1,
\;
W_{\text{boundary}} \odot x_1
\right),
\end{equation}
where $\odot$ denotes element-wise multiplication. This formulation assigns the highest penalty to tumor interior pixels closest to tumor boundaries, encouraging the model to preserve sharp contrast transitions associated with blood-brain barrier disruption \cite{heye2014assessment}, while gradually relaxing supervision toward the interior. The overall training objective is given by,
\begin{equation}
\mathcal{L} =
\mathcal{L}_{\text{latent}}
+ \lambda_{\text{pixel}} \mathcal{L}_{\text{pixel}}
+ \lambda_{\text{boundary}} \mathcal{L}_{\text{boundary}}.
\end{equation}
\subsubsection{Inference.}
At inference time, only a non-contrast MRI image $x_0$ is required. It is encoded in the latent space as $z_0 = \mathcal{E}(x_0)$, which initializes the latent trajectory at $t=0$. Note that during inference, we do not need to provide a tumor mask as input.  The model predicts the transported latent directly using Eq.\eqref{eq:predict_latent}. This is typically done in under four inference steps, effectively saving the computation required for iterative process in diffusion models. The terminal latent $\hat{z}_1$ is finally decoded using the VAE decoder to obtain the synthesized contrast-enhanced image $\hat{x}_1 = \mathcal{D}(\hat{z}_1)$.




\section{Experiments}

\noindent\textbf{Datasets:} The \textbf{BraTS 2023 GLI} \cite{li2024brain} dataset consists of 1{,}470 Brain MRI volumes. All scans are resampled to an isotropic resolution of $1\,\mathrm{mm}^3$ and spatially cropped to $240\times 240\times155$. The data was split into $1001/250/219$ MRI volumes for training/validation/testing respectively. The \textbf{In-House} dataset in this study consists of 720 paired Liver MRI images. These are T1-weighted volumetric images, acquired using a free-breathing radial gradient-echo sequence on a 1.5 T MRI scanner. Pre-contrast and hepatobiliary phase images were obtained following intravenous administration of a hepatobiliary contrast agent. We used this dataset for both Zero-shot Inference and Finetuning results, to assess the model’s generalization ability. The data was split into 600 and 120 MRI images for training and testing respectively.

\noindent\textbf{Implementation:}
Axial slices are retrieved from the MRI volumes that are first normalized by clipping intensity values between the 0.1th and 99.9th percentiles and rescaled to $[0,1]$. They are then zero-padded to $256\times256$ for model training. A U-Net \cite{ronneberger2015u} is used to predict drift $v_{\theta}$, with its weights initialized to the weights of pre-trained text-to-image model SDXL \cite{podell2023sdxl}. We use timestep distribution $\pi(t)$ of 4 discrete timesteps, each with equal probability. For both pixel and boundary-aware loss, we use $\ell_1$ loss with $\lambda_{pixel} = 18$ and $\lambda_{boundary} = 14$. An $\ell_2$ loss is used for the latent loss. The strength of tumor bias in TuBAM is set to $\alpha_{tumor} = 0.5$. We set noise parameter $\sigma = 0.008$. Model was trained for 50{,}000 steps on a single A100 GPU, and the optimization is performed using AdamW with a learning rate of $4 \times 10^{-5}$ and a batch size of $8$. Training takes approximately 10 hours and inference takes on an average 0.097 seconds per image.

\subsection{Results}

Our TuLaBM was compared with Pix2Pix~\cite{isola2017image}, 
ResViT~\cite{dalmaz2022resvit}, Palette~\cite{saharia2022palette}, 
I$^2$SB~\cite{liu20232} and D$^3$M~\cite{pang2025d}. The quantitative 
metrics for evaluating the generation results are Structural Similarity 
Index (SSIM\%) and Peak Signal-to-Noise Ratio (PSNR).

\noindent\textbf{BraSyn (Brain MRI).}
As reported in Table~\ref{tab:results}, TuLaBM achieves a whole-image 
SSIM of $92.40$\% and tumor-region SSIM of $88.66$\%, outperforming all other baselines, 
demonstrating that TuBAM and the boundary-aware loss effectively 
enhance synthesis fidelity in clinically critical tumor regions.

\noindent\textbf{Cleveland Clinic Liver MRI.}
To evaluate generalization across anatomies, we assess TuLaBM on an 
in-house liver MRI dataset (collected at Cleveland Clinic) in both zero-shot and fine-tuned settings. 
In the zero-shot setting, TuLaBM achieves a PSNR of $16.57$\,dB and 
an SSIM of $58.30$\%, outperforming all other baselines without any domain-specific training. After fine-tuning on 5 liver MRI volumes, TuLaBM further improves to 
$19.34$\,dB PSNR and $63.60$\% SSIM, 
demonstrating strong cross-anatomy adaptability, confirming that TuLaBM's latent formulation generalizes effectively to unseen anatomical domains.

\noindent\textbf{Qualitative Results.}
Fig.~\ref{fig:qualitative} shows representative synthesized CE-MRI 
images from both datasets. TuLaBM produces sharper tumor boundaries 
and more faithful contrast enhancement compared to all baselines, 
closely matching the ground truth in both brain and liver anatomies, 
while Pix2Pix and ResViT exhibit false positive enhancement, Palette 
suffers from blurring artifacts, and D$^3$M shows less precise tumor 
boundary delineation.

\noindent\textbf{Inference Efficiency.}
As shown in Table~\ref{tab:results}, TuLaBM achieves inference in 
$0.097$ seconds per image using only four sampling steps, matching 
the speed of GAN-based methods (Pix2Pix: $0.024$\,s, ResViT: 
$0.028$\,s) and being orders of magnitude faster than other diffusion-based 
methods (I$^2$SB: $11.8$\,s, D$^3$M: $4.58$\,s), making it 
practically viable for clinical deployment.

\noindent\textbf{Ablation Study.}
Table~\ref{tab:ablation} reports the contribution of each proposed 
component on the BraSyn dataset. Removing the boundary-aware loss 
(TuLaBM w/o BL) leads to a notable drop in both whole-image SSIM 
($92.40 \rightarrow 88.13$\%) and tumor-region SSIM 
($88.66 \rightarrow 84.17$\%), confirming its role in sharpening 
tumor interfaces. Further removing TuBAM (TuLaBM w/o BL and TuBAM) 
results in additional degradation across all metrics, demonstrating 
that TuBAM independently contributes to improved tumor-region 
synthesis by amplifying intra-tumor feature aggregation during 
latent bridge evolution.

\begin{table}[t]
\centering
\caption{Quantitative comparison on the BraTS 2023 GLI (BraSyn) and the In-house Dataset. The means and standard deviations of the PSNR and SSIM between generated and real images, as well as the results within the tumor regions.
}
\label{tab:results}
\resizebox{\textwidth}{!}{%
\begin{tabular}{lccccccccc}
\toprule
 & \multicolumn{4}{c}{BraSyn} & \multicolumn{4}{c}{Cleveland Clinic (Liver MRI-in house)} &  \\
\cmidrule(lr){2-5} \cmidrule(lr){6-9}
 & \multicolumn{2}{c}{Whole Image} & \multicolumn{2}{c}{Tumor Region} & \multicolumn{2}{c}{Zero-shot} & \multicolumn{2}{c}{Finetuned} &  \\
\cmidrule(lr){2-3} \cmidrule(lr){4-5} \cmidrule(lr){6-7} \cmidrule(lr){8-9}
Model & PSNR & SSIM & PSNR & SSIM & PSNR & SSIM & PSNR & SSIM & Inference Time (s) \\
\midrule
Pix2Pix & $23.53_{\pm2.89}$ & $87.70_{\pm4.25}$ & $15.31_{\pm4.40}$ & $65.37_{\pm18.38}$ & $16.46_{\pm0.93}$ & $48.84_{\pm0.08}$ & $18.65_{\pm0.39}$ & $50.62_{\pm0.06}$ & 0.0241 \\
ResViT  & $23.54_{\pm2.81}$ & $87.43_{\pm4.19}$ & $15.21_{\pm4.41}$ & $65.19_{\pm17.87}$ & $16.21_{\pm1.67}$ & $53.73_{\pm0.10}$ & $18.28_{\pm0.95}$ & $56.15_{\pm0.07}$ & 0.0279 \\
Palette & $18.95_{\pm5.02}$ & $39.97_{\pm29.09}$ & $14.00_{\pm4.75}$ & $67.86_{\pm17.10}$ & $15.11_{\pm1.90}$ & $45.03_{\pm2.18}$ & $17.21_{\pm0.29}$ & $52.13_{\pm0.06}$ & 1.22 \\
I$^2$SB & $25.01_{\pm2.94}$ & $89.11_{\pm3.99}$ & $17.05_{\pm4.81}$ & $71.59_{\pm16.66}$ & $16.47_{\pm1.70}$ & $50.14_{\pm0.40}$ & $18.70_{\pm1.29}$ & $56.17_{\pm0.01}$ & 11.8 \\
D$^3$M  & $25.11_{\pm3.33}$ & $90.95_{\pm3.86}$ & $17.33_{\pm4.56}$ & $73.21_{\pm16.22}$ & $16.50_{\pm3.45}$ & $57.13_{\pm1.02}$ & $18.73_{\pm1.40}$ & $58.19_{\pm0.033}$ & 4.58 \\
\midrule
\rowcolor{oursrow}
TuLaBM & $\mathbf{25.16}_{\pm2.58}$ & $\mathbf{92.40}_{\pm1.51}$ & $\mathbf{18.34}_{\pm3.01}$ & $\mathbf{88.66}_{\pm5.78}$ & $\mathbf{16.57}_{\pm0.55}$ & $\mathbf{58.30}_{\pm9.17}$ & $\mathbf{19.34}_{\pm1.53}$ & $\mathbf{63.60}_{\pm9.45}$ & \textbf{0.097} \\
\bottomrule
\end{tabular}%
}
\end{table}

\begin{table}[t]
\tiny
\centering
\caption{Ablation study on the BraTS2023-GLI (BraSyn) dataset 
investigating the individual contribution of the Boundary-Aware Loss 
(BL) and Tumor-Biased Attention Mechanism (TuBAM) in TuLaBM. Mean 
and standard deviation of PSNR (dB) and SSIM (\%) are reported for 
whole-image and tumor-region evaluation. 
}
\label{tab:ablation}
\begin{tabular}{lcccc}
\toprule
 & \multicolumn{2}{c}{Whole Image} & \multicolumn{2}{c}{Tumor Region} \\
\cmidrule(lr){2-3} \cmidrule(lr){4-5}
Model & PSNR & SSIM  & PSNR & SSIM  \\
\midrule
\rowcolor{oursrow} TuLaBM& $\mathbf{25.16}_{\pm2.58}$& $\mathbf{92.40}_{\pm1.51}$& $\mathbf{18.34}_{\pm3.01}$& $\mathbf{88.66}_{\pm5.78}$\\
TuLaBM w/o BL& $22.53_{\pm2.46}$& $88.13_{\pm1.35}$& $16.34_{\pm3.22}$& $84.17_{\pm7.20}$\\
TuLaBM w/o BL and TuBAM& $22.19_{\pm2.44}$& $88.09_{\pm5.78.34}$& $16.38_{\pm3.18}$& $84.16_{\pm7.13}$\\
\bottomrule
\end{tabular}
\end{table}

\section*{Conclusion}
In this paper, we presented TuLaBM, a latent bridge matching framework 
for efficient CE-MRI synthesis from NC-MRI. By operating in a learned 
latent space and introducing the Tumor-Biased Attention Mechanism 
(TuBAM) together with a boundary-aware loss, TuLaBM achieves superior 
tumor region synthesis fidelity, with a notable gain in tumor-region 
SSIM ($88.66$ vs.\ $73.21$ for D$^3$M), while reducing inference to 
under $0.097$ seconds per image. Experiments on BraTS2023-GLI and an 
in-house liver MRI dataset further confirm its strong generalization 
across anatomies in both zero-shot and fine-tuned settings, 
establishing TuLaBM as an accurate and clinically efficient approach 
to contrast-free MRI enhancement.

%
%
\bibliographystyle{splncs04}
\bibliography{references}
\end{document}